\newcount\subeqnno
\def\bse{\begingroup
\refstepcounter{equation}
\subeqnno=\arabic{equation}
\setcounter{equation}{0}
\def\theequation{\the\subeqnno\alph{equation}}}
\def\ese{\setcounter{equation}{\the\subeqnno}\endgroup}

\newcommand{\mb}[1]{\ifmmode#1\else\mbox{$#1$}\fi}

\newcommand\ga{\mb{\gamma}}
\newcommand\de{\mb{\delta}}

\newcommand\la{\mb{\lambda}}

\newcommand\si{\mb{\sigma}}

\newcommand\De{\mb{\Delta}}

\newcommand\Ph{\mb{\Phi}}

\newcommand\calC{\mb{{\cal C}}}
\newcommand\calD{\mb{{\cal D}}}

\newcommand\calG{\mb{{\cal G}}}
\newcommand\calH{\mb{{\cal H}}}
\newcommand\calI{\mb{{\cal I}}}
\newcommand\calJ{\mb{{\cal J}}}

\newcommand\calL{\mb{{\cal L}}}
\newcommand\calM{\mb{{\cal M}}}
\newcommand\calN{\mb{{\cal N}}}

\newcommand{\beq}{\begin{equation}}
\newcommand{\eeq}{\end{equation}}
\newcommand{\nn}{\nonumber}
\newcommand{\bea}{\begin{eqnarray}}
\newcommand{\eea}{\end{eqnarray}}
\newcommand{\norm}[1]{\parallel \! {#1} \! \parallel}

\newcommand{\inprod}[2]{\langle {#1}, {#2} \rangle}
\newcommand{\inproda}[2]{\{ {#1}, {#2} \}}

\newcommand{\deriv}[2]{\frac{d {#1}}{d {#2}}}

\newcommand{\x}{\mb{\times}}

\newcommand{\Ad}{\mb{{\rm Ad}}}

\newcommand{\tr}{\mb{{\rm tr}}}

\newcommand{\ul}[1]{{\underline{#1}}}

\documentstyle[12pt]{article} 
\topmargin = - 0.5 cm
\begin{document}
\bibliographystyle{unsrt}

\newtheorem{result}{Result}
\newtheorem{definition}{Definition}



\begin{flushright}
DAMTP-1999-140
\end{flushright}
\vspace*{0.5cm}
\begin{center}
\LARGE{Vacuum Geometry} \\
\vspace*{0.7cm}
\large{
Nathan\  F.\  Lepora
\footnote{e-mail: N.F.Lepora@damtp.cam.ac.uk} }\\
\vspace*{0.2cm}
{\small \em Department of Applied Maths and Theoretical Physics,\\
Cambridge University, England.\\ }
\vspace*{0.2cm}

{October 1999}
\end{center}
\vspace*{0.2cm}
\begin{abstract}
We analyse symmetry breaking in general gauge theories paying
particular attention to the underlying geometry of the theory. In this
context we find two natural metrics upon the vacuum manifold: a  
Euclidean metric associated with the scalar sector, and another
generally inequivalent metric associated with the gauge
sector. Physically, the interplay between these metrics gives rise to
many of the non-perturbative features of symmetry breaking.  
\end{abstract}
							
\thispagestyle{empty}
\newpage
\setcounter{page}{1}


\section{Introduction}

Non-Abelian gauge theories are the modern setting for theories of
particle interaction. Both the strong and electroweak
interactions are described by such theories. For electroweak
interactions the full gauge symmetry is not apparant at low energies,
with it assumed to be broken in some high energy
phase transition. This same concept of symmetry breaking may be taken
further to unify the strong and electroweak interactions. 

Central to the symmetry breaking scheme is the concept of a
vacuum. The symmetry is broken by the coupling between this vacuum and
the original gauge fields. This coupling induces mass for the gauge
fields associated with the broken symmetries, whilst
the gauge fields associated with the residual symmetries do not
couple, remain massless and form the residual gauge theory.

However, gauge symmetry breaking has further implications. Their
structure implies the existence of 
non-perturbative effects, which imply specific physical consequences
for the theory under consideration. Examples of non-perturbative
effects include sphaleron processes, topological vortices,
topological monopoles and dynamically stable vortices. Their spectrum
and properties are very model dependant, depending upon the
particular pattern of symmetry breaking.


In this paper we are concerned with the specific geometric
structure of the vacuum manifold and how it relates to the spectrum
and properties of many non-perturbative effects. 
Our approach is to associate two natural homogenous metrics on the
vacuum manifold. One metric associated with the structure of the
scalar sector, the other associated with the gauge sector. Comparison
of the relative geodesic structures determines many of the
non-perturbative features of the theory. 

It transpires that the boundary conditions for many non-perturbative
solutions depends upon the comparative geodesic structure of the scalar
and gauge metrics. Embedded vortices correspond to 
mutually geodesic circles, whilst embedded monopoles correspond to
mutually geodesic two-spheres and sphalerons correspond to mutually
geodesic three spheres. Furthermore, this approach also relates
the Aharanov-Bohm scattering of embedded vortices to the holonomy of
geodesics with respect to the gauge sector metric.

One should note that the material in this paper stems from a similar
geometrical analysis of the Weinberg-Salam model~\cite{metomb}. For
illustration we perform the analysis here in parallel to that. 

The plan of this paper is as follows. We firstly review gauge
symmetry breaking using a notation compatible with the latter sections
of this paper. We then discuss the associated gauge and scalar
metrics. Finally we apply this discussion to the determination of
non-perturbative solutions and their properties in general gauge
theories. 

\section{Gauge symmetry Breaking}

Consider when a gauge symmetry $G$ is broken to a
residual symmetry $H$. A Lagrangian $\calL$ describes the interaction of
the gauge fields $A^\mu$ with scalar fields $\Phi$. These
gauge fields $A^\mu$ take values in the Lie algebra $\calG$, whilst
the scalar fields take values in the vector space $V$. Minimisation of
this Lagrangian yields a vacuum with gauge invariance under only the
subgroup $H$.

The gauge symmetry $G$ has an action upon the scalar fields, described
by the representation $D$. Correspondingly the Lie 
algebra also has a derived action upon the scalar fields, described by
the derived representation $d$ such that $D(e^X) = e^{d(X)}$ for all
$X \in \calG$. 

The interaction of the scalar-gauge sector is then represented by the
Lagrangian 
\beq
\label{lag}
\calL = -\frac{1}{4} \inprod{F^{\mu \nu}}{F_{\mu \nu}} +
\calD^\mu \Phi^\dagger\: \calD_\mu \Phi
- V[\Phi],
\eeq
with field tensor 
\beq
F^{\mu \nu} = \partial^\mu A^\nu - \partial^\nu A^\mu + [A^\mu,
A^\nu],
\eeq
covariant derivative 
\beq
\calD^\mu = \partial^\mu + d(A^\mu),\\
\eeq 
and $\inprod{\cdot}{\cdot}$ the inner product on $\calG$.

There is some freedom in the choice of non-degenerate inner product
$\inprod{\cdot}{\cdot}$ on $\calG$, constrained to
be invariant under the adjoint action $\Ad(G)$. We parameterise the
possible inner products by splitting the Lie algebra $\calG$ into its
commuting 
subalgebras $\calG = \calG_1 \oplus \cdots \oplus \calG_n$ and then
defining the inner product $\inprod{\cdot}{\cdot}$ to consist of a
linear combination of the inner products on each $\calG_l$.
The parameters describing the linear combination of inner products
then specify the relevant length scales between the commuting
components of $\calG$. We shall show that these are related to the
gauge coupling constants.

Explicitly the general $\Ad(G)$-invariant inner product on $\calG$ is
\beq
\label{inprod}
\inprod{\sum_{l} X_l}{\sum_{l} Y_l} = 
\sum_{l} {s_l}\inproda{X_l}{Y_l}_l, \ \ \ \ X_l, Y_l \in \calG_l
\eeq
with real, positive parameters $\{s_1, \cdots, s_n\}$ and
\bse
\bea
\inproda{X_l}{Y_l}_l &=& -p\: {\rm Re}[{\rm tr}(X_l Y_l)],
\hspace{6em} \calG_l \ {\rm simple}\\
\inproda{X_l}{X_l}_l &=& 1,\ \ \ \exp(2\pi X_l)=1,
\hspace{4.5em} \calG_l \cong u(1).
\eea
\ese
Taking an orthonormal basis $\{X^1_l,\cdots,X^{\dim(\calG_l)}_l\}$
for $\calG_l$ with respect to $\inproda{\cdot}{\cdot}_l$, the unit norm
generators with respect to $\inprod{\cdot}{\cdot}$ are
\beq
\norm{q_l(s_1, \cdots s_n) X^i_l} = 1 \ \ \ \ {\rm (no\ sum)}.
\eeq
With these generators the covariant derivative explicitly takes the form
\beq
\calD^\mu = \partial^\mu + \sum_l q_l A^{i\mu}_l X^i_l,
\eeq 
from which the scales $\{q_1,\cdots,q_n\}$ are interpreted as the
corresponding gauge coupling constants.

Symmetry breaking is seen through minimisation of the
Lagrangian~(\ref{lag}), one solution of which is the vacuum 
\beq
\label{vac1}
\Phi(x)=\Phi_0,\ \ \ \ A^\mu=0,
\eeq
for some arbitrary $\Phi_0$ minimising $V[\Phi]$. The other minima
of Eq.~(\ref{lag}) collectively give rise to the vacuum manifold of
degenerate equivalent solutions   
\beq
M = D(G)\Phi_0,
\eeq
a homogenous manifold contained in $V$.  

Around the vacuum in Eq.~(\ref{vac1}) the residual field theory has
gauge symmetry $H$. The effect of the interaction between this vacuum
and the gauge fields is to give mass to those gauge fields
not associated with the generators of $H$. However the
basis $\{X^i_l\}$ does not generally correspond to a basis associated
with mass eigenstates. To find this basis of mass eigenstates one must
consider an orthogonal transformation of the basis vectors
$\{X^i_l\}$. 

We shall use a gauge equivalence argument to find this basis of
mass eigenstates. Firstly we split the gauge bosons into massless and
massive families associated with the decomposition
\beq
\label{reddec}
\calG = \calH \oplus \calM.
\eeq
This has the algebraic property
\beq
\Ad(H) \calH \subseteq \calH,\ \ \ \ \Ad(H)\calM \subseteq \calM,
\eeq
which physically represents that the massive gauge bosons
are not equivalent to the massless gauge bosons under $H$.

Massless gauge bosons are generated by elements of $\calH$. These
split into gauge eigenstate families corresponding to the
decomposition
\beq
\calH = \calH_1 \oplus \cdots \oplus \calH_m,
\eeq
of $\calH$ into commuting subalgebras. The space of zero mass
eigenstates is then  
\beq
\label{massless}
\sum_j e_j Y^i_j B^{i\mu}_j,
\eeq
with $\{e_j Y^1_j, \cdots, e_j Y^{\dim(\calH_j)}_j\}$ forming an
orthonormal basis for each $\calH_j$ and $e_j$ the corresponding
scale. The massless gauge fields are denoted by $B^{i\mu}_j$.

Massive gauge bosons are generated by elements of $\calM$. This
splits into mass-degenerate families corresponding to the
decomposition of 
\beq
\label{mdec}
\calM = \calM_1 \oplus \cdots \oplus \calM_l,
\eeq
into $\Ad(H)_G$ irreducible subspaces of $\calM$. Physically the
$H$-equivalent gauge bosons are degenerate in mass, which split the
spectrum of $H$-equivalent gauge boson generators into mass equivalent
families. The space of massive eigenstates is then
\beq
\label{massive}
\sum_f g_f Z^i_f W^{i\mu}_f,
\eeq
with $\{g_f Z^1_f, \cdots, g_f Z^{\dim(\calM_f)}_f\}$ forming an
orthonormal basis for $\calM_f$, with scale $g_f$ and $W^{i\mu}_f$ the
corresponding massive gauge fields.

Taken together Eqs.~(\ref{massless}) and (\ref{massive}) specify the
covariant derivative to be  
\beq
\calD^\mu = \partial^\mu + \sum_j e_j Y^i_j B^{i\mu}_j +
\sum_f g_f Z^i_f W^{i\mu}_f,
\eeq 
explicitly composed of the residual and broken parts.
Again the scales $\{e_j\}$ and $\{g_f\}$ are interpreted as the
associated gauge coupling constants. They depend
upon the scales $\{q_i\}$, with the explicit dependance determined by
the orthogonal transformation that relates the $\{X_i\}$ basis of
$\calG$ to the mass eigenstate basis above. 

\section{Vacuum Geometry}

For both the scalar and gauge sectors we now explicitly calculate
their associated metrics. We also calculate the associated
isometry and isotropy groups for each, and from these groups specify the
corresponding geodesic structures.

Having obtained two inequivalent homogenous metrics on the vacuum
manifold we then compare their structure. We relate their isometry and
isotropy groups, and determine when curves are mutually geodesic with
respect to both metrics.

\subsection{Gauge Sector}

The main structure associated with the gauge sector is the inner
product $\inprod{\cdot}{\cdot}$ of Eq.~(\ref{inprod}). It specifies
several important related features of the gauge sector. First of all
it defines the gauge kinetic term in Eq.~(\ref{lag}), introducing the
gauge coupling constants as the relative scales. Secondly it
stipulates the massive and massless gauge boson
generators to be the orthogonal subspaces $\calM$
and $\calH$ of Eq.~(\ref{reddec}). Finally it renders
the degenerate mass eigenstate families $\calM_i$ to be mutually
orthogonal.   

We shall use this inner-product to define the gauge-sector
metric. The definition is achieved by associating the massive
generators $\calM$ with tangent spaces to the vacuum manifold in a
natural way. Then the inner product $\inprod{\cdot}{\cdot}$ on $\calM$
induces a homogenous metric on the vacuum manifold. We find 
the corresponding isometry group of this metric to be
the gauge group $G$ and the isotropy group to be the
residual symmetry $H$. 

Explicitly, observe that the tangent space to $M$ at $\Phi_0$ may be
expressed 
\beq
T_{\Phi_0}M = d(\calM) \Phi_0.
\eeq
More generally the corresponding tangent space at $\Phi=g \Phi_0 \in
M$ is, for any $g \in G$, 
\beq
\label{tgt-iso}
T_\Phi M = D(g)T_{\Phi_0} M = d(\Ad(g)\calM) \Phi, 
\eeq
Transitivity over $M$ guarantees a natural isomorphism between any 
tangent space and some $\Ad(g) \calM$.

Using the isomorphism implied by Eq.~(\ref{tgt-iso}), the inner
product $\inprod{\cdot}{\cdot}$ associates a corresponding
metric on $M$
\beq
\label{metric-gauge}
h(X_1 \Phi, X_2 \Phi)_\Phi = \inprod{X_1}{X_2},\ \ \  X_1, X_2 \in
\Ad(g)\calM. 
\eeq
The precise form is parameterised by the scales $\{g_f\}$ of
Eq.~(\ref{massive}).  

This metric has the gauge symmetry group $G$ of isometries
\beq
h(D(g)T_1,D(g)T_2)_{D(g)\Phi}=h(T_1, T_2)_{\Phi},\ \ \ \ 
g \in G. 
\eeq
More precisely, by Eq.~(\ref{tgt-iso}) the action of $g \in G$ upon
$h(\cdot,\cdot)$ is 
\beq
h(D(g)T_1,D(g)T_2)_{D(g)\Phi} = \inprod{\Ad(g)X_1}{\Ad(g)X_2} =
\inprod{X_1}{X_2} = h(T_1,T_2)_{\Phi}.
\eeq
The above isometries represent the maximal group acting upon $\calG$
that leaves $\inprod{\cdot}{\cdot}$ invariant.

The isotropy group of this isometry group at the point
$\Phi_0$ in $M$ is the subgroup $H$ such that 
\beq
D(H) \Phi_0 = \Phi_0,
\eeq
which gives the isomorphism
\beq
\label{iso-gauge}
M \cong \frac{G}{H}.
\eeq 
Thus we recover the familiar relation for the vacuum manifold, but
now explicitly associated with the gauge sector metrical structure.

Given the above isotropy and isometry properties of the gauge metric
we can use the isomorphism~(\ref{iso-gauge}) to calculate the 
corresponding geodesics upon $M$. This
geodesic structure follows from some results of differential
geometry. Specifically, these results
examine the geodesic structure on the coset space, but may
be simply carried back to the vacuum manifold to give the results
that we require. The relevant results are quoted here, although the
full approach is described in detail by Kobayashi and
Nomizu~\cite[chapter X]{Nomi2}. 

The gauge sector geodesic structure is simply
\begin{quote}\em
the geodesics on $M$ with respect to the metric $h(\cdot,\cdot)$, passing
through $\Phi_0$ are: 
\beq
\label{geo-gauge}
\ga_X=\{\exp(Xt)\Phi_0: t \in {\bf R}\},
\eeq
with $X \in \calM$.
\end{quote}

\subsection{Scalar Sector}

The structure associated with the scalar sector is a vector space of
scalar field values $V$ equipped with a real Euclidean inner
product Re$[\Psi^\dagger \Phi]$. Regarding the vacuum manifold as
embedded within the vector space of scalar field values,
a natural metric may be induced on $M$ by specifying its form on each
tangent space to be that of the Euclidean inner product. 
 
Explicitly, regard the tangent space to $M$ at $\Phi \in M$, specified
by Eq.~(\ref{tgt-iso}), as a vector subspace of $V$
\beq
T_\Phi M \subset V.
\eeq
Then a corresponding metric is induced from the real Euclidean inner
product  
\beq
\label{metric-scalar}
g(T_1,T_2)_\Phi = {\rm Re}[T_1^\dagger T_2], \ \ \ T_1, T_2 \in
T_\Phi M. 
\eeq

This metric has a group $I$ of isometries acting on $V$ by the
$\De$-representation  
\beq
g(\De(a)T_1,\De(a)T_2)_{\De(a)\Phi}=g(T_1, T_2)_{\Phi},\ \ \ \ 
a \in I. 
\eeq
consisting of those elements of ${\rm GL}(V)$ that leave the metric
invariant. Since $I \subset {\rm GL}(V)$ the representation of $I$
upon $V$ is induced by
\beq
\De(I) \subset f({\rm GL}(V)),
\eeq
with $f$ the fundamental representation of ${\rm GL}(V)$ upon $V$.

The isotropy group of $I$ upon $M$ at the point $\Phi_0$ is the
subgroup $J$ such that  
\beq
\De(J) \Phi_0 = \Phi_0,
\eeq
This gives the isomorphism
\beq
\label{iso-scalar}
M \cong \frac{I}{J},
\eeq
representing a second isomorphism with a coset space, and separate from
the usual one associated with the gauge sector in
Eq.~(\ref{iso-gauge}). 

The scalar coset space isomorphism found above in
Eq.~(\ref{iso-scalar}) is explicitly related to the
scalar sector structure. In general the gauge group $G$ is a subgroup
of the isotropy group $I$, with the representations coinciding so that
$D(G)\subseteq \De(I)$. Symmetries not contained in $G$ are
interpreted as global symmetries of the vacuum manifold not 
apparent in the full gauge theory.

We now prove $D(G)\subseteq\De(I)$. This follows from the gauge
invariance of the Lagrangian~(\ref{lag}), so that any $g \in G$ 
preserves the Euclidean norm,
\beq
{\rm Re}[\{D(g)T\}^\dagger D(g)T\}] =  {\rm Re}[T^\dagger T],
\eeq
for all $T$ in any $T_\Phi M$. Thus $D(g)^\dagger D(g)=1$, implying 
\bse
\bea
g(D(g)T_1,D(g)T_2)_{a\Phi} &=& {\rm Re}[\{D(g)T_1\}^\dagger D(g)T_2]\\
&=& g(T_1, T_2)_{\Phi}.
\eea
\ese
Proving our statement. 

As for the gauge sector the importance of 
isomorphism~(\ref{iso-scalar}) is the calculational use of the
isotropy and isometry properties to determine the associated geodesic
structure upon $M$. Since our
method is the same for both the scalar and gauge sectors this
allows direct comparison to be drawn between them. 

Associated with the isomorphism~(\ref{iso-scalar}) is the decomposition
\beq
\label{redI}
\calI = \calJ \oplus \calN,
\eeq
where $\calN$ is associated with the tangent space to $M$ at $\Phi_0$
\beq
T_{\Phi_0} M = \de(\calN) \Phi_0,
\eeq
with $\de$ the derived representation of $\De$. The relevant inner
product, which defines the decomposition~(\ref{redI}), is defined by:
\begin{quote}\em
associated with the $I$-invariant metric $g(\cdot, \cdot)$ upon $M$ is
an inner product upon $\calN$
\beq
\inproda{X}{Y} = g_{\Phi_0}(X\Phi_0, Y\Phi_0) = {\rm
Re}[\{X\Phi_0\}^\dagger Y\Phi_0]; 
\eeq
extendable to a unique inner product on $\calI$ such that
$\inproda{\calJ}{\calN}=0$. 
\end{quote}

Geodesic structure then follows analogously to Eq.~(\ref{geo-gauge}):
\begin{quote}\em
the geodesics on $M$ with respect to the scalar
metric~$g(\cdot,\cdot)$, passing through $\Phi_0$ are: 
\beq
\label{geo-scalar}
\ga_X=\{\exp(Xt)\Phi_0: t \in {\bf R}\},
\eeq
for elements $X \in \calN$.
\end{quote}

One should note that this geodesic structure
does not in general coincide with that of the gauge sector,
although it may do for certain scales $\{q_i\}$. The geodesic structure
on $M$ with respect to $G/H$ may be interpreted as homogenously
squashed with respect to that from $I/J$. More exactly we have shown
that $G$ is a subgroup of the isometry group $I$, hence the associated
gauge metric has correspondingly less invariance. 

For instance $S^3$ admits a family of $SU(2) \x U(1)$ invariant metrics
that continuously deforms to an $SU(2) \x SU(2)$ invariant metric;
and $S^7$ admits a family of $SO(5) \x SU(2)$ invariant metrics
that continuously deform to the $SO(8)$ invariant metric. 

\subsection{Scalar-Gauge Geometry}
\label{sec-scga}

In summary, there are two inequivalent metrics on the electroweak
vacuum manifold 
associated with the scalar and gauge sectors. We shall now determine
how the structure of these metrics relate to each other. 
Comparing the respective symmetry groups determines those symmetries
that are shared. These shared symmetries define 
submanifolds whose geodesics are mutually geodesic with respect to both
metrics. 

The scalar and gauge metrics, $g(\cdot,\cdot)$ and $h(\cdot,\cdot)$,
have the following isometry group decompositions with 
respect to their isotropy groups 
\bse
\bea
\calI &=& \calJ \oplus \calN,\\
\calG &=& \calH \oplus \calM,
\eea
\ese
where their group structures are related by
\bse
\bea
H &\subset& J,\\
G &\subset& I,\\
\calM &\subset& \calI,
\eea
\ese
and the representations of $G$ and $I$ are coincident on the
intersection
\beq
D(G) \subseteq \De(I).
\eeq
Also the tangent space to $M$ at $\Phi_0$ is related to $\calM$ and
$\calN$ by 
\beq
T_{\Phi_0} M = d(\calM) \Phi_0 = \de(\calN)\Phi_0.
\eeq

It is important to understand how the metrics $g(\cdot,\cdot)$ and
$h(\cdot,\cdot)$ are related. It transpires that this is through
the decomposition of $\calM$ into its $\Ad(H)_G$ irreducible subspaces
\beq
\label{mdecom}
\calM = \calM_1 \oplus \cdots \oplus \calM_k,
\eeq
which defines a decomposition of the tangent space into its $D(H)$
irreducible subspaces $T^i_{\Phi_0}M = d(\calM_i)\Phi_0$
\beq
\label{tdecom}
T_{\Phi_0}M = T_{\Phi_0}^1 M \oplus \cdots \oplus T_{\Phi_0}^k M.
\eeq
Then at $\Phi_0 \in M$ the metrics $g(\cdot,\cdot)$ and
$h(\cdot,\cdot)$ are related by a bilinear transformation of
$T_{\Phi_0}M$ taking the particular form
\bse
\bea
\label{relmet}
g(\sum_i X_i \Phi_0, \sum_j Y_j \Phi_0) = \sum_{ij} \la_i \la_j
h(X_i \Phi_0, Y_j \Phi_0),\\
\label{relmetb}
{\rm where}\ \ \ \la_i = \frac{g(X_i \Phi_0, X_i \Phi_0)}{h(X_i
\Phi_0, X_i \Phi_0)},
\eea
\ese
and $X_i \Phi_0 \in T_{\Phi_0}^iM$.

Result~(\ref{relmet}) is established by noting that since both metrics
are bilinear they are related by an element $A \in GL(T_{\Phi_0}M)$
\beq
\label{bilin}
g(T_1, T_2)_{\Phi_0} = h(AT_1, AT_2)_{\Phi_0},\ \ \ 
A \in GL(T_{\Phi_0}M).
\eeq
Now observe $D(H) \subseteq GL(T_{\Phi_0}M)$, is the {\em maximal}
subgroup of $GL(T_{\Phi_0}M)$ under which both metrics are
invariant. Then $D(h)A=AD(h)$ for all $h \in H$. It follows that the
eigenspaces of $A$ are the irreducible spaces of 
$T_{\Phi_0}M$ under $D(H)$, from which one associates
Eqs.~(\ref{mdecom}, \ref{tdecom}) and establishes the
result~(\ref{relmet}, \ref{relmetb}).     
This may be easily generalised to all $\Phi \in M$ by considering the
action of $G$ on Eq.~(\ref{relmet}).

The decomposition~(\ref{tdecom}) also describes the geodesic structure
of $M$ with respect to $g(\cdot,\cdot)$ and $h(\cdot,\cdot)$ in a
rather nice way. Applying Eqs.~(\ref{geo-scalar}) and
(\ref{geo-gauge}), the submanifolds 
\beq
\label{subM}
M^i = D(\exp(\calM_i))\Phi_0,\ \ \ T_{\Phi_0}M^i = T^i_{\Phi_0} M 
\eeq
are the {\em maximal} totally geodesic submanifolds of $M$ with respect
to {\em both} metrics $g(\cdot,\cdot)$ and $h(\cdot,\cdot)$. They
naturally decompose $M$ into totally geodesic component parts.

One should note that the $\Ad(H)_G$-irreducible subspaces
of $\calM$ also relate to the mass eigenstates of the massive gauge
bosons. It is interesting how the same decomposition arises in two, at
first sight, apparently unrelated areas of the gauge theory; however
the association is that both are governed by the action of $H$ within
the broken theory.

\section{Physical Implications}

In summary of the previous section: there are two homogenous metrics on
the vacuum manifold associated with the scalar and gauge sectors of
the scalar-gauge theory. These two sectors induce metrics upon the
vacuum manifold associated with the symmetries of their
respective sector. Generally the geodesics with respect to these two
metrics coincide on totally geodesic submanifolds. These totally
geodesic submanifolds are embedded in $M$ such that at any point their
collection of tangent spaces do not intersect and collectively from the
tangent space to $M$ at that point.

Given this structure we examine its relation to the
spectrum of non-perturbative solitonic-type solutions present for
a general symmetry breaking. It transpires that the embedded vortices
correspond to the mutually geodesic paths, whilst 
monopoles correspond to mutually geodesic two-spheres and sphalerons
to mutually geodesics three-spheres. This approach also interprets the
scattering of embedded vortices in terms of the holonomy of their
respective geodesics. Furthermore the dynamical stability of a semi-local
vortices, in their respective semi-local limit, is seen to correspond
to an extreme limit of the gauge metric.

\subsection{Embedded Vortices}

Embedded vortices correspond to Nielsen-Olesen vortices embedded in
a general symmetry breaking~\cite{Vach94}. As such their boundary
conditions define circular paths on the vacuum manifold. Thus it
might be expected that their spectrum and properties should correspond
to the geometry of the vacuum manifold. This is what we find. Their
boundary conditions correspond to the paths that are mutually geodesic
with respect to {\em both} metrics.

Formally, an embedded vortex is defined by the embedding
\bea
G \ \ \ &\rightarrow& H\nn \\
\cup \ \ \ &\ & \cup\\
U(1) &\rightarrow& 1,\nn
\eea
with the general Ansatz 
\bse
\bea
\label{emb1}
\Phi(r, \theta) &=& f_{\rm NO}(r) e^{X\theta}\Phi_0, \\
\label{emb2}
\ul{A}(r,\theta) &=& \frac{g_{\rm NO}(r)}{r} X \ul{\hat{\theta}},
\eea
\ese
where $X\in \calG$ is the vortex generator. 
One may consider only $X \in \calM$, since these minimise the
magnetic energy~\cite{me2}. Thus one considers only
Ans\"atze with boundary conditions geodesic with respect to
the gauge metric $h(\cdot, \cdot)$ .

The above vortex Ansatz is a solution provided that~\cite{Barr92}\\
(i) The scalar field must be single-valued. Hence the boundary
conditions describe a closed geodesic with 
$e^{2 \pi X}\Phi_0=\Phi_0$.\\ 
(ii) The Ansatz is a solution to the equations of motion; then fields
in the vortex do not induce currents perpendicular (in Lie algebra
space) to it~\cite{Vach94}. This may be equivalently phrased
as~\cite{me2}:   
$X$ is a vortex generator if Re$[(X\Phi_0)^\dagger X^\perp \Phi_0] =
0$ for all $X^\perp$ such that $\inprod{X^\perp}{X}=0$.

Condition (ii) can be conveniently restated in terms of the
corresponding metrics:  
\begin{quote}\em 
$X$ is a vortex generator if the associated tangent vector
$T=X\Phi_0$ satisfies $g(T, T^\perp)=0$ for all $T^\perp$ such that
$h(T, T^\perp)=0$. 
\end{quote}
Referring to the discussion around Eq.~(\ref{tdecom}), we see that $T$
must lie in one of the $D(H)$-irreducible subspaces of $T_{\Phi_0}M$
that relate the two metrics. Namely $X$ is an element of any of the
$\Ad(H)_G$-irreducible subspaces of $\calM$ in the decomposition  
\beq
\calM = \calM_1 \oplus \cdots \oplus \calM_n.
\eeq

It is interesting that the geodesics defined from the
$\Ad(H)$-irreducible subspaces $\calM_i$ are the only geodesics which
are {\em simultaneously geodesic with respect to both metrics}. From a
mathematical point of view this is because these geodesics define
submanifolds of the vacuum manifold with coincident scalar and
gauge metrics (to an overall factor). From a physical
point of view, this may be interpreted as a minimisation of both the
scalar and gauge sectors of the action integral.

Vortices within the same family $\calM_i$ have the same stability
properties because of their gauge equivalence. For topologically
stable embedded vortices the associated geodesic $\ga_X(t) =
D(\exp(Xt)\Phi_0$ has a conserved topological charge corresponding to
the element of $\pi_1(G/H)$ it represents. Dynamically
stable vortices are discussed next.

\subsection{Dynamical Stability}

Embedded vortices may become dynamically stable if there
exists a limit of the coupling constants $\{q_i\} \rightarrow
\{\tilde{q}_i\}$ where the symmetry breaking takes the
form~\cite{Pres} 
\beq
\label{sl}
G_g \x U(1)_l \rightarrow H,\ \ \ {\rm with}\ 
H \cap U(1)_l = {\bf 1}.
\eeq
In this limit there exists for small scalar potentials dynamically
stable semi-local vortices~\cite{vach91}, and by continuity their
stability persists for nearby values of the coupling constants. Here
we show that this is related to an extreme limit of the gauge
sector metric. In this limit it becomes only well defined on a circle 
corresponding to the boundary conditions of the semi-local vortex.

Algebraically Eq.~(\ref{sl}) is related to the decomposition of
$\calM$ into its $\Ad(H)_G$ irreducible subspaces~\cite{me2}:
\begin{quote}\em
when $\calM_l$ represents a non-trivial projection of $\calC$, the
centre of $\calG$, onto $\calM$ then that family may define
dynamically stable vortices.
\end{quote}
Such $\calM_l$'s are one-dimensional, with their vortices invariant
under $H$. 

This qualitative difference of the gauge theory when the coupling
constants take values $\{\tilde{q}_i\}$ is also apparent in the form
of the gauge sector metric $h(\cdot,\cdot)$. When the coupling
constants take the value $\{\tilde{q}_i\}$,
the inner product $\inprod{\cdot}{\cdot}$ becomes ill defined; being
well defined only on the subalgebra $u(1)_l$, where it takes
the form 
\beq
\label{ip1}
\inprod{X}{Y}_{q_l} = - \frac{1}{q_l^2}\tr X \tr Y.
\eeq
Referring to the discussion around Eqs.~(\ref{metric-gauge}) and
(\ref{subM}), we see that the metric is only defined upon the
one-dimensional submanifold $M_l=D(U(1)_l)\Phi_0 \subset M$, where it
takes the form 
\beq
h_{q_l}(T_1,T_2)_\Phi = \inprod{X_1}{X_2}_{q_l},\ \ \ T_i = d(X_i)
\Phi \in T_\Phi M_l, 
\eeq 
with $a \in U(1)_l$, relating $\Phi=D(a) \Phi_0$. 

The metric $h(\cdot, \cdot)$ thus picks out a prefered submanifold
$M_l \subset M$ over which it is well defined. Physically, this
submanifold corresponds to those points that may be reached by a gauge
transformation from $\Phi_0$. Other points within the vacuum manifold
may only be reached by a global transformation. It is precisely such
submanifolds that define the boundary conditions for vortices that
may be dynamically stable.

\subsection{Combination Electroweak Vortices}

Recall that the two metrics are related by
\bse
\bea
g(\sum_i X_i \Phi_0, \sum_j Y_j \Phi_0) = \sum_{ij} \la_i \la_j
h(X_i \Phi_0, Y_j \Phi_0),\\
{\rm where}\ \ \ \la_i = \frac{g(X_i \Phi_0, X_i \Phi_0)}{h(X_i
\Phi_0, X_i \Phi_0)},
\eea
\ese
which is the reason for claiming the coincidence of the metrics on
certain submanifolds defined by the $\Ad(H)_G$ irreducible subspaces
$\calM_i$. In sec.~(\ref{sec-scga}) we then showed that these
$\calM_i$'s contain the vortex generators, since to define solutions
the scalar boundary conditions must be mutually geodesic with respect
to both metrics. 

The constants $\{\la_i\}$, which define the relative scales of the two
metrics, are dependant upon the gauge coupling constants $\{q_i\}$.
Should coupling constants take critical values such
that $\la_i = \la_j$, the two metrics $g(\cdot, \cdot)$ and $h(\cdot,
\cdot)$ become coincident (to a factor) on a larger submanifold
$M_{ij}=D(\exp(\calM_i \oplus \calM_j)\Phi_0$, containing both $M_i$
and $M_j$. Then one concludes that the vortex generators are members
of a larger family $\calM_i \oplus \calM_j$, and one refers to these
extra vortices as {\em combination} vortices --- interpreting them as
two vortices in the original families combined together.

For critical coupling constants such that $\la_i=\la_j$ one of two
situations may occur, depending upon the details of the geometry:\\
(i) The isometry group of the gauge sector metric $h(\cdot, \cdot)$
enlarges. Then vortices in the two families $\calM_i$ and $\calM_j$
becomes equivalent under global transformations of these newly
acquired symmetries.\\
(ii) The isometry group of the gauge sector metric $h(\cdot, \cdot)$
stays the same. The equivalence of vortices remains the same as the
non-critical case; however, the solution set of embedded vortices
increases to include the combination vortices.

These two situations relate to the symmetry properties of
scattering two vortices. For two different vortices
heading towards each other the symmetry of the scattering process
causes them to travel along the line joining the centres both before
and after they meet. However, when the vortices are the same, as
happens in (i) above, they may scatter at right angles to the original
path, with the natural motion passing through a toroidal charge two
configuration. This charge two configuration corresponds to a
combination vortex. 

Although the combination vortex solutions are not solutions for
non-critical coupling constants, it may be possible that static
deformed solutions persist. Such solutions would consist
of perturbed solutions around the embedded vortex, with the
perturbations determined by substitution into the field equations.  

\subsection{Non-Abelian Aharanov-Bohm Scattering}

Associated with the magnetic flux of a vortex is the Wilson line
integral   
\beq
\label{wilson}
U(\theta) = P \exp \left( \int_{0}^{\theta} {\bf A} \cdot d{\bf l}
\right) \subset G, 
\eeq
at infinite radius. Then $U(\theta)$ dictates the parallel transport
of matter fields around a vortex, with fermionic fields $\Psi$ 
transported to $U(2\pi)\cdot\Psi$ around a vortex. By diagonalisation 
one then associates components $\Psi_i$ with phase shifts
$\xi_i$. 

The Wilson line integral is fundamental to understanding the
interaction of a vortex with charged matter
fields~\cite{Wilz}. Non-trivial fermionic components $\Psi_i$ interact
with the vortex by an Aharanov-Bohm cross section 
\beq
\deriv{\si}{\theta} = \frac{1}{2\pi k}
\frac{\sin^2(\xi_i/2)}{\sin^2(\theta/2)},
\eeq
whilst trivial components $e^{2 \pi \xi_k}=1$ interact by an Everett
cross section~\cite{Ever}. 

Substitution of Eq.~(\ref{emb2}) into Eq.~(\ref{wilson}) gives the
Wilson line integral for an embedded vortex, with generator $X \in
\calM$, 
\beq
U(\theta) = \exp(X\theta) \subset G.
\eeq
Vortex boundary conditions restrict $U(2\pi) \in H$. Also, by 
Eq.~(\ref{emb1}) the asymptotic scalar field satisfies
$\Phi_\infty(\theta)= D(U(\theta))\Phi_0$. For a given fermion
representation this specifies completely the scattering of a fermion
component off an embedded vortex.

In \cite{Wilz} the Wilson line integral is related to holonomy,
referring to non-trivial parallel transport around the vortex. In fact,
we now show that this holonomy refers to precisely the {\em parallel
transport with respect to the gauge sector metric around its closed
geodesics}. 

Parallel transport with respect to the gauge sector metric $h(\cdot,
\cdot)$ on $M$ is related to the symmetry groups $G$ and $H$
by~\cite{Nomi2}: 
\begin{quote}\em
the parallel transport of tangent vector $T \in T_{\Phi_0}M$ along the
geodesic $\ga_X(t)=D(\exp(Xt))\Phi_0$ to the point $\ga_X(s)$ is 
\beq
T' = D(\exp(Xs))T.
\eeq
\end{quote}
The holonomy is non-trivial when $\ga_X(s)=\ga_X(0)$ and $T' \neq
T$. 

Hence geodesic parallel transport and the
Wilson line integral are related. For a tangent vector $T \in
T_{\Phi_0} M$, its parallel transport to $\ga_X(\theta)$ is
\beq
\label{du}
T'= D(U(\theta))T.
\eeq
Interpreting tangent vectors as components of the scalar field we see
that Eq.~(\ref{du}) also refers to a global transformation of $G$ upon
the scalar field. The same global transformation is relevant
to the parallel transport of a fermion representation along the
geodesic. Thus we see that the Wilson line integral refers to
parallel transport along a geodesic with respect to the gauge
sector metric $h(\cdot, \cdot)$. One should recall also that the
geodesics relevant to vortices are geodesic with respect to both
metrics, however it is only the gauge sector holonomy that
is relevant to scattering.

\subsection{Embedded Monopoles}

Embedded, or fundamental, monopole solutions are t'Hooft-Polyakov
monopoles embedded within the gauge theory. To avoid ambiguity between
embedded $SU(2)$ or embedded $SO(3)$ monopoles, the embedding is
described in terms of the algebras
\bea
\calG &\rightarrow& \calH  \\
\cup &\ & \cup \\
su(2) &\rightarrow& u(1),
\eea
with the general Ansatz
\bse
\bea
\underline{\Ph}(\underline{r}) &=& f_{\rm mon}(r) \underline{\hat{r}},\\
A^\mu_a(\underline{r}) &=& \frac{g_{\rm mon}(r)}{r}
\epsilon^\mu_{ab} X_b, 
\eea
\ese
where $f_{\rm mon}(r)$ and $g_{\rm mon}(r)$ are the usual monopole
profile functions. Notationally, we treat $\Ph$ a
vector within ${\bf R} \cdot D(SU(2))\Phi_0$ and use
an orthonormal basis $\{X_1, X_2, X_3\}$ for $su(2)$, with $X_3 \in
\calH$. One may consider only $X_1, X_2 \in \calM$ since these
minimise the magnetic energy. Thus one considers only
Ans\"atze with boundary conditions geodesic two-spheres with respect
to the $h(\cdot, \cdot)$ metric.

By relating the above embedding to the spectrum of embedded vortices,
one arrives at the following result~\cite{me2}:
\begin{quote}\em
embedded monopoles are defined by two generators $X_1, X_2$
such that 
\beq
X_1, X_2 \in \calM_i,\ {\rm with}\  [X_1, X_2] \in \calH.
\eeq
\end{quote}
Hence the gauge equivalence classes of embedded monopoles are similar
to those of embedded vortices, but with extra constraints.

Boundary conditions for the embedded monopole define a two-dimensional
spherical submanifold $D(SU(2))\Phi_0 \subset M$. Analogous to
embedded vortices these submanifolds are {\em simultaneously geodesic
with respect to both metrics}. However, in contrast to vortices, there
may be other geodesic two-spheres not associated with embedded
monopoles: for instance the Sphaleron configurations, which we
discuss next. 

\subsection{Embedded Sphalerons}

Analogously to embedded vortices and monopole one may embed 
sphalerons~\cite{Mant} in a general symmetry breaking. To avoid
ambiguity between 
$SU(2)$ and $SO(3)$ Sphalerons the embedding is discussed in terms of
the algebras. For an $su(2)$ embedded Sphalerons the embedding is
\bea
\calG &\rightarrow& \calH  \\
\cup &\ & \cup \\
su(2) &\rightarrow& {\bf 1};
\eea
whilst for an $su(2) \oplus u(1)$ Sphaleron the embedding is
\bea
\calG &\rightarrow& \calH  \\
\cup &\ & \cup \\
su(2)\oplus u(1) &\rightarrow& u(1)_{\rm diag}.
\eea
The corresponding Ansatz\"e are simply the usual ones embedded in a
larger theory.

By a similar analysis to that of monopoles one may conclude that an
embedded $su(2)$ sphaleron corresponds to simultaneously geodesic
three-spheres; whilst an embedded $su(2) \oplus u(1)$ sphaleron
corresponds to a geodesic three-sphere submanifold with respect to
$g(\cdot, \cdot)$ and a squashed three-sphere submanifold with respect
to $h(\cdot, \cdot)$. Contained within this embedded three-sphere is
a two sphere that is simultaneously geodesic with respect to both
metrics, and has no topology associated with it.

Physically, embedded sphalerons relate to the vacuum structure of the
gauge theory. Inequivalent vacua of the quantum field theory are
labelled by the elements of $\pi_3(G/H)$ and correspond to the
Chern-Simons number of the vacua. Sphalerons represent the midpoint of
a sequence of configurations passing from one vacuum to another; they
are thus physically relevant when they have an embedding associated
with a non-trivial element of $\pi_3(G/H)$.

\section{Discussion}

In this final section we briefly discuss some extra points of note,
and indicate some possible extensions to the work of this paper.
\\
(i) {\bf Simplicity of Electroweak Theory}\\
We showed that the maximal symmetry group of any metric upon the
vacuum manifold
is the isometry group $I$, with the gauge group $G\subseteq I$. For
Electroweak theory $I=SU(2)_I \x SU(2)_K$ and $G=SU(2)_I 
\x U(1)_Y$. In fact this constitutes the {\em smallest} dimensional
example of such a structure~\cite{metomb}.\\
(ii) {\bf Energetics and Curvature}\\
The metrical structure determines an associated curvature of the
vacuum manifold. It seems natural that the energy of embedded
vortices should correspond with the sum of the relevant sectional
curvatures associated with the scalar and gauge sectors on the
particular submanifold of $M$ associated 
with the scalar boundary conditions of the vortex. Coefficients of
this sum would naturally be related to those of the scalar potential,
and the value of this sum to the stability of the vortex.\\
(iii) {\bf Non-Abelian Zero Modes}\\
The zero modes of a configuration correspond to zero energy
deformations of that configuration. Since the relevant configurations 
correspond to geodesic submanifolds, the zero modes must preserve
this property. Thus zero modes should be related to 
the metrical structure of the vacuum manifold. \\
(iv) {\bf Mathematics of the Scalar Metric}\\
Often the scalar sector metric, associated with $I/J$, corresponds
to a symmetric space. Consequently the scalar sector holonomy is
trivial. It would be interesting to determine exactly when a symmetric
space is associated, and whether this is always required to give a
trivial holonomy to the scalar sector. 

\bigskip


{\noindent{\Large{\bf Acknowledgements.}}}
\nopagebreak
\bigskip
\nopagebreak

I acknowledge Kings College, Cambridge for a Junior Research
Fellowship and thank Conor Houghton, Tom Kibble and Paul Saffin for
interesting conversations related to this work.  This work was
supported in part by the European Commission under the Human Capital
and Mobility program, contract no. CHRX-CT94-0423.

\bigskip



\end{document}